\documentstyle[aps,prl,epsfig]{revtex}
\begin{document}
\twocolumn

\title{Evolution of Ultracold, Neutral Plasmas }
\author{ S. Mazevet, L. A. Collins, J. D. Kress}
\address{ Theoretical Division, Los Alamos National Laboratory, Los Alamos, NM 87545 }
\date{submitted to PRL}
\maketitle
\begin{abstract}
We present the first large-scale simulations of an ultracold,
neutral plasma, produced by photoionization of laser-cooled xenon
atoms, from creation to initial expansion, using classical molecular
dynamics methods with open boundary conditions. We reproduce many of
the experimental findings such as the trapping efficiency of electrons
with increased ion number, a minimum electron temperature achieved on
approach to the photoionization threshold, and recombination into
Rydberg states of anomalously-low principal quantum number.  In
addition, many of these effects establish themselves very early in the
plasma evolution ($\sim$ ns) before present experimental observations
begin.

\end{abstract}

That a common characteristic connects such diverse environments as the
surface of a neutron star, the initial compression stage of an inertial
confinement fusion capsule, the interaction region of a high-intensity
laser with atomic clusters, and a very cold, dilute, and
partially-ionized gas within an atomic trap, seems at first rather
remarkable. Yet all these cases embrace a regime known as a
strongly-coupled plasma. For such a plasma, the interactions among the
various constituents dominate the thermal motions. The plasma coupling
constant $\Gamma_{\alpha}$, the ratio of the average electrostatic
potential energy Z$_{\alpha}$/a$_{\alpha}$ [ Z$_{\alpha}$, the
charge; a$_\alpha$, the ion-sphere radius = $3$/($4\pi n_{\alpha}{^{1/3}})$;
and $n_{\alpha}$, the number density for a given component $\alpha$] to
the kinetic energy k$_B$T$_{\alpha}$, provides an intrinsic measure of
this effect \cite{ichimaru}. When $\Gamma_{\alpha}$ exceeds unity,
various strong-coupling effects commence such as collective modes and
phase transitions. For multi-component plasmas, the coupling constants
need not be equal or even comparable, leading to a medium that may
contain both strongly- and weakly- coupled constituents.

Since temperatures usually start around a few hundred Kelvin, most
plasmas found in nature or engineered in the laboratory attain
strongly-coupled status from high densities, as in the case of a
planetary interior or a shock-compressed fluid\cite{nellis,nova}, or
from highly-charged states, as in colloidal or ``dusty'' plasmas
\cite{dusty}. In both situations, the particle density usually rises
well above 10$^{18}$/cm$^{3}$.  On the other hand, ion trapping and
cooling methods have produced dilute, strongly-coupled plasmas by
radically lowering the temperature. At first, these efforts were
limited to nonneutral plasmas confined by a magnetic
field\cite{dubin}.  However, recently, new techniques
\cite{tom1,tom2,tom3,gall1} have generated neutral, ultracold plasmas,
free of any  external fields at densities of the order of
10$^8$-10$^{10}$/cm$^{3}$ and temperatures at microkelvins ($\mu$K).

Two methods, one direct and one indirect, but both employing laser
excitation of a highly-cooled sample of neutral atoms, have
successfully created such neutral plasmas. The direct approach employs
the photoionization of laser-cooled xenon atoms \cite{tom1,tom2,tom3},
while the indirect generates a cold Rydberg gas in which a few
collisions with room-temperature atoms produces the
ionization\cite{gall1,uconn}.  In both cases, the electron and ion
temperatures start very far apart with the former from 1K to 1000K and
the latter remaining at the initial atom temperature of a few $\mu$K.
This implies that the coupling constant for the electrons ($\Gamma_e$)
ranges between 1 and 10, while the ions begin with $\Gamma_i \sim
1000$.  Based on the results presented below, the following picture
emerges. As the system evolves, the electrons reach a quasi-equilibrium
while at the same time beginning to move the ions through long-range
Coulomb interactions.  Eventually, the ions also heat, and the whole
cloud begins to expand with the ions and electrons approaching
comparable temperatures. The former processes occur on the order of
pico- to nano- seconds, while the full expansion becomes noticeable on a
microsecond scale.  Therefore, by following the progress of the system,
we can study not only the basic properties of a strongly-coupled plasma
but also its evolution through a variety of stages. The path to
equilibration still remains a poorly understood process for
strongly-coupled plasmas in general.

The dilute character of the ultracold plasmas provides a unique
opportunity to explore the intricate nature of atomic-scale,
strongly-coupled systems, both through real-time imaging of the sample
to study  wave phenomena, collective modes, and even phase transitions,
and through laser probes to examine internal processes. The initial
experiments have already discovered unexpected phenomena such as more
rapid than expected ion expansion\cite{tom2} and Rydberg populations
that strongly deviate from those predicted from standard electron-ion
recombination rates \cite{tom3,seaton,hanh97,stevef75}.  So
far, the experiments have examined the later expansion stage of the
whole plasma cloud, which occurs on the scale of microseconds. However,
the nature of the plasma at this stage depends strongly on its
evolution from its creation. While ultra-fast laser techniques offer
the eventual prospect of probing these early times, for now, only
simulations can provide an understanding of the full evolution of these
systems. We focus on the first such generated plasma\cite{tom1} from
the photoionization of cold Xenon atoms; however, the findings also
provide insight into other ultracold systems.

Since the electron temperature greatly exceeds the Fermi temperature
for this system, we may effectively employ a two-component classical
molecular dynamics (MD) approach in which the electrons and ions are
treated as distinguishable particles, interacting through a modified
Coulomb potential. We have employed several different forms from an
{\it{ab-initio}} Xe$^{+} + e^{-}$ pseudopotential\cite{TM} to a simple
effective cut-off potential\cite{HM} based on the de Broglie thermal
wavelengths; we found little sensitivity of the basic properties to
this choice. This finding stems from the dominance of the interactions
by the very long-range Coulomb tails since the average particle
separations are of the order of $10^4$\AA  (1 to 2 times the de Broglie wavelength for electrons at a temperature of 0.1K).  The short-range
modification of the Coulomb potential basically prevents particles with
opposite charges from collapsing into a singularity. Due to the open
and nonperiodic boundary conditions, sample size effects can play a
critical role in the simulations\cite{russ}. Therefore, we must employ
particle numbers of the basic order of the experiments, requiring
efficient procedures for handling the long-range forces. In addition,
the extended simulation times needed to model the entire plasma
evolution [fs to $\mu$s] demand effective temporal treatments. To this
end, we have employed multipole-tree procedures\cite{tree} in
conjunction with reversible reference system propagator
algorithms\cite{respa} (r-RESPA) through the parallel 3D MD program
NAMD\cite{namd}, generally used in biophysical applications. These
procedures allow us to treat samples of between 10$^2$ and 10$^4$
particles [N = N$_i$/2 = N$_e$/2]; most simulations were performed with
N = 10$^3$.

To reproduce the initial experimental conditions\cite{tom1}, we
distribute the ions randomly according to a 3D Gaussian, whose rms
radius, $\sigma$, matches the desired {\em{ion}} number density
n$_{\rm{i}}$ [$\rm{n}_i=N_i/(2\pi\sigma^2)^{3/2}$] and impose a
Maxwell-Boltzmann velocity distribution at $T_i=1\mu$K. To each ion, we
associate an electron placed randomly in an orbit of radius $r_{o}$=
50\AA.  The electron velocities point in random directions but with
fixed magnitude, determined by the photoionization condition [K$_{e}^{i}
\equiv \frac{1}{2}m_ev_e^2 = \hbar\omega - 1/r_{o}$, with $\omega$, the
laser frequency and $v_e^2$ as a sum over all three spatial components].
This idealized prescription models the photoionization process by allowing 
each electron to escape the Coulomb well with a final prescribed energy. By
varying the laser frequency, we can control the final effective
electron temperature, just as in the experiments. We
tested this particular ionization model against simulations where both
the initial electron-ion radius r$_{o}$ and the form of the ion
distribution are varied.  Overall, we find little sensitivity to these
variations on the basic initial conditions. Finally, we take a
kinematic definition of temperature as the average kinetic energy per
particle T$_{\alpha}$ = $\frac{1}{3N_{\alpha}k_B} \sum_i m_{\alpha}
v_{\alpha,i}^{2}$, where the sum runs over all particles i of type
$\alpha$.

The plasma evolves through several stages. In the first or
photoionization stage, which lasts on the order of fs, the electrons
climb out of the Coulomb well and become basically freely-moving
particles. In the next stage, the electrons reach a quasi-equilibrium
at T$_e$ due to their fast intra-particle collisions.  Following this
stage, the electron-ion collisions begin the slow process of heating
the ions, which can require up to $\mu$s.
This process can also be viewed in terms of an electron pressure term
in a hydrodynamical formulation\cite{tom2}. Then, the whole
cloud of ions and electrons begins a systematic expansion. 
This progression clearly evinces itself in Fig. 1, which shows the temporal
behavior of electron and ion average kinetic energies, using an effective 
mass for the ion of m$_i$ = 0.01 amu in order to accelerate the evolutionary
process. The averaged kinetic energy of the electrons and ion become 
comparable and the cloud perceptibly expands in about 20 ns. Scaling this 
value by $\sqrt{\frac{m_{Xe}}{m_i}}$ yields an estimate for this time in a Xe
plasma of about 1$\mu$s, in line with the experiments\cite{tom1}. Also 
consistent with the experimental measurements\cite{tom2}, these preliminary
calculations indicate that a large fraction of the kinetic energy 
transferred by the electrons results in outward translational motion 
for the ions. This implies that the kinetic temperature defined above
 coincides with the usual themodynamic temperature (random thermal motion) 
for the electrons only. In all other simulations discussed, the ions carry 
the mass of Xe.

\vspace{.4cm}\begin{figure}[t]
\begin{center}
\epsfig{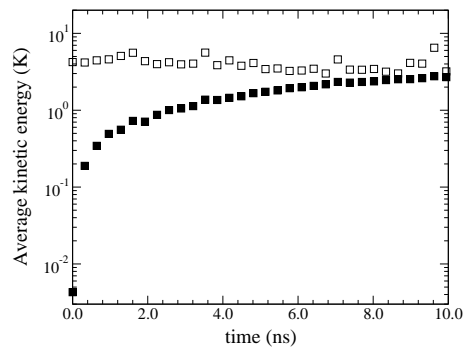}
\vspace{.3cm}\caption{ Schematic variation of the electron (open square)
and ion (filled square) average kinetic energy per particle as a function of time for a model plasma with m$_i$=0.01 amu.}
\end{center}
\end{figure}

Since we shall use the simulations to understand intermediate stages in
the development of the ultracold plasma, we need to establish the
validity of the procedure by comparing to experiments. Two initial
observations have particular significance: 1) the number of trapped
electrons rises with the number of ions created for a fixed final
temperature, and 2) the electrons attain a certain minimum temperature
no matter how small $\hbar\omega$ becomes. Figure 2a shows that the
number of trapped electrons as a function of the number of ions and
initial electron temperature basically follows the general experimental
trends\cite{tom1}. Some of the electrons, freed by the photoionization
process from ions with sufficient energy, escape the atomic cloud and
never return, leaving an overall positively charged system. This
residual charge then effectively traps the remaining electrons so that
the center of the distribution resembles a neutral plasma. The larger
the number of ions produced, the more effective the confinement of the
electrons.

In Fig.2b, we present for n$_{\rm{i}}$=4.32x10$^9$ ions/cm$^{3}$, the
electron temperature T$_{e}^{f}$ after the initial equilibration in the
simulations as a function of the excess initial temperature T$_{e}^{i}
[= \frac{2}{3} K_{e}^{i}$] given to the electrons during the
photoionization process.  The plateau at about 5K [$\Gamma_e \sim 1$]
for small T$_{e}^{i}$ appears quite pronounced.  This effect arises due
to a mechanism usually designated as ``continuum lowering''\cite{ichimaru}. Despite
the small energies and enormous distances involved, at least by usual
atomic standards, the interaction of an ion with its neighbors still
has a noticeable effect due to the very long-range of the Coulomb
potential.  This shifts the appropriate zero of energy from the
isolated atom to the whole system. Using a simple binary interaction,
we can estimate this energy difference for an electron halfway between
the ions [$\sim 2/a_{ion}$] at around 3K for this density, in
qualitative agreement with the simulations.  Even for a photon energy
near the atomic threshold, the electron within the plasma will still
gain this minimum energy. This same effect also explains the
dependence of the final electron temperature on the ion density.  In
this situation, an increase in the ion density enhances the zero of
energy shift and leads to an increase in the final electron
temperature.

\vspace{.5cm}\begin{figure}[t]
\begin{center}
\epsfig{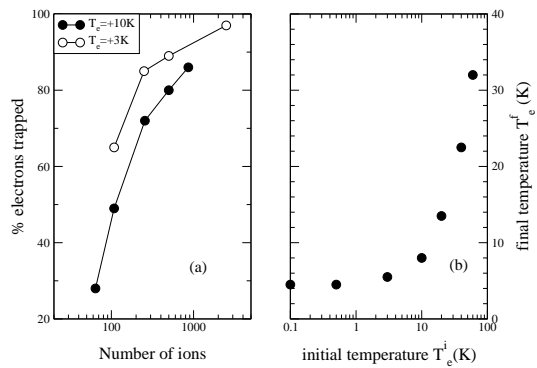}
\vspace{.3cm}\caption{(a) Percentage of electrons trapped as a function of the number of ions created for two electron temperatures (3K and 10K). (b) Variation of the final electron temperature T$_{e}^{f}$ as a function of the initial excess kinetic energy represented an associated temperature T$_{e}^{i}$.}
\end{center}
\end{figure}

The molecular dynamics simulations yield additional properties of the
ultracold plasma.  We paid particular attention
to the electron plasma frequency $f_e$ since experiments\cite{tom2}
have used this quantity as an indirect means of deducing the density of
the ultracold plasma. Given the nature of the system with open boundary
conditions and quasi-equilibration among the electrons, the question
arises as to whether $f_e$ has a precise definition. Neglecting 
temperature dependence, the electron plasma frequency is given by
\begin{equation}
\label{pf}
f_e=\frac{1}{2\pi}\sqrt{\frac{e^2n_e}{m_e}},
\end{equation}
where e is the elementary charge, $n_e$ the electron density, and $m_e$
the electron mass. From the MD simulations, the
electron plasma frequency is obtained from the Fourier
transform of the electron velocity autocorrelation
function\cite{hansen}. We find, reassuringly, that a distinct though
broadened peak arises near the $f_e$ predicted by Eq.\ref{pf} in the
regime in which the electrons reach a quasi-equilibration. For
example, at a density of 4.32x10$^{10}$ ions/cm$^{3}$, an electron
temperature of 3K, and simulation time extending up to 18ns, MD
gives plasma frequencies of 1.2GHz (1.4 GHz) while Eq.\ref{pf} yields 1.57GHz (1.87GHz) for trapped electrons at
70\% (100\%) of the ion density.  When the ion
density is decreased to a value of 4.32x10$^{9}$ ions/cm$^3$ at the
same electron temperature, $f_e$  from the MD simulations becomes
0.5Ghz, also in accordance with Eq.\ref{pf} . In general, we
find that the number of particles used in the simulation cell has
little effect on the determination of the electron plasma frequency
beyond providing a better statistical sample.

\begin{figure}[t]
\begin{center}
\epsfig{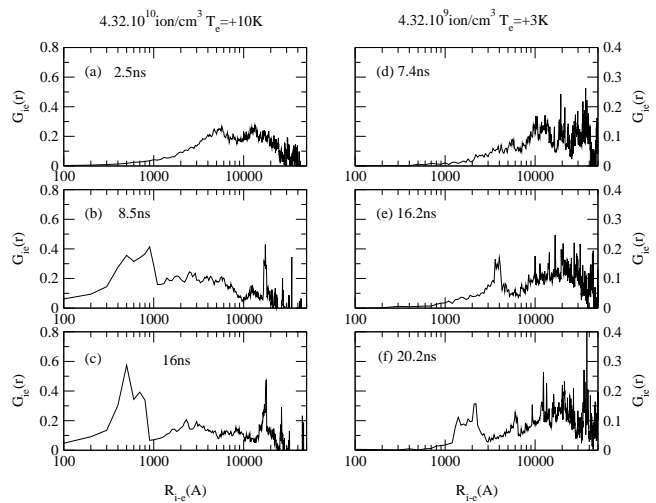}
\vspace{.2cm}
\caption{Time evolution of the modified ion-electron pair correlation
function $G_{ie}$ as a function of density and electron temperature:
(a)-(c)  4.32x10$^{10}$ ions/cm$^3$ and 10K, and (d)-(f) 4.32x10$^{9}$
ions/cm$^3$ and 3K.}
\end{center}
\end{figure}

We now turn to the nature of the constituents as the plasma evolves and
pay particular attention to the formation of Rydberg atoms as noted in
the experiments\cite{tom3,gall1,uconn}. To this end, we examine over
space and time the ion-electron pair correlation function $g_{ie}$(r),
which gives the probability of finding an electron a distance r from a
reference ion\cite{HM}. To enhance the examination of the closer
encounters, we have employed a modified form $G_{ie}$(r) by subtracting
a uniform distribution and averaging $g_{ie}$(r) over a small time
interval of 1.5 ns. Figure 3 displays $G_{ie}$(r) for two
representative cases. The open boundary conditions, which force the
distribution to zero at some large distance, give $G_{ie}$(r) a
distinctly different behavior than seen in a fluid or periodic system.

Figures 3(a)-(c) display the development of a plasma for
n$_i$=4.32x10$^{10}$ ions/cm$^3$ and T$_e \sim$ 10K in the regime in
which the electrons have reached quasi-equilibration and the ions have
started to heat. After 2.5ns, the electron distribution departs from
its initial form, especially beyond a radius of 10000\AA.  By 8.5ns, we
begin to observe an increase in $G_{ie}$ just below 1000\AA. This
signifies the formation of Rydberg states in an average principal
quantum number n $\sim$ 40, a conclusion supported by an examination of
movies of the simulation. These movies also show stable Rydberg atom
configurations over many orbital periods of the electrons. A simple
model\cite{stevef75} that balances various collisional processes, using
strictly atomic cross sections, predicts a much larger value (n $\geq$
100). For conditions closer to the experimental measurements (lower
density), as depicted in Figs.  3(d)-(f), we find a significant Rydberg
population after only 20ns (Fig.3f).  The electron distribution shifts
to around 2000\AA, corresponding to a principal quantum number n $\sim$
60. In both cases, the Rydberg population is estimated to be between
5\% and 10 \% of the total number of electrons. These findings closely
resemble those of the experiments\cite{tom3}, though for later times
($\sim \mu$s) and indicate that collective or density effects may play
an important role by changing the accessible Rydberg-level distribution
or cross sections through long-range interactions. By examining the
electron mean square displacement over time, we can identify two stages
in the evolution of the cold ionized gas. First, after the ionization
of the atoms, the electrons diffuse throughout the system for several
nanoseconds with no significant Rydberg atom formation. Second, the
electrons reach the edge of the cloud and begin systematic multiple
traverses of the system. The Rydberg atom population becomes noticeable
only several nanoseconds into this second stage.  The feature that
attracts the most interest is the relatively short time scale required
to establish the neutral plasma and produce a noticeable Rydberg
population.

In summary, we have followed the evolution of an ultracold, neutral
plasma over a broad range of temporal stages with classical molecular
dynamics simulation techniques.  We find general agreement with
experimental observations of the number of trapped electrons, the
minimum of electron temperature, and the production of Rydberg atoms in
low-lying states. The latter two conditions especially demonstrate the
importance of strong-coupling or density effects on the basic atomic
interactions.  In addition, we have found that the electron plasma
frequency appears as a valid tool to probe the state of the system.
Important to the understanding of the temporal development of these
plasmas, we discovered that recombination and the formation of
long-lived Rydberg states occur rapidly on the order of nanoseconds. Our
studies continue in an effort to push into the fully expanded regime
by using hydrodynamical methods tied to the molecular
dynamics.

We wish to acknowledge useful discussions with Dr. S. Rolston (NIST),
Prof. T. Killian (Rice), and Prof. P. Gould (U. of Connecticut).  We
thank Dr. N. Troullier for providing the Xe pseudopotential. Work
performed under the auspices of the U.S. Department of Energy (contract
W-7405-ENG-36) through the Theoretical Division at the Los Alamos
National Laboratory.

\end{document}